\documentclass[nohyperref]{article}

\usepackage{microtype}
\usepackage{graphicx}
\usepackage{subfigure}
\usepackage{booktabs} % for professional tables
\usepackage{enumitem}

\usepackage{hyperref}

\usepackage[accepted]{icml2022}

\usepackage{amsmath}
\usepackage{amssymb}
\usepackage{mathtools}
\usepackage{amsthm}
\usepackage{commands}

\usepackage[capitalize,noabbrev]{cleveref}

\theoremstyle{plain}

\theoremstyle{definition}

\theoremstyle{remark}

\usepackage[textsize=tiny]{todonotes}

\icmltitlerunning{Challenges and Opportunities of Shapley values in a Clinical Context}

\begin{document}

\twocolumn[
\icmltitle{Challenges and Opportunities of Shapley values in a Clinical Context}

\icmlsetsymbol{equal}{*}

\begin{icmlauthorlist}
\icmlauthor{Lucile Ter-Minassian}{equal,yyy}
\icmlauthor{Sahra Ghalebikesabi}{equal,yyy}
\icmlauthor{Karla Diaz-Ordaz}{comp,sch}
\icmlauthor{Chris Holmes}{yyy,sch}
\end{icmlauthorlist}

\icmlaffiliation{yyy}{University of Oxford}
\icmlaffiliation{comp}{The London School of Hygiene \& Tropical Medicine}
\icmlaffiliation{sch}{The Alan Turing Institute}

\icmlcorrespondingauthor{Lucile Ter-Minassian}{lucile.ter-minassian@spc.ox.ac.uk}

\icmlkeywords{Interpretable ML, Healthcare}

\vskip 0.3in
]
\printAffiliationsAndNotice{\icmlEqualContribution} % otherwise use the standard text.

\begin{abstract}
With the adoption of machine learning-based solutions in routine clinical practice, the need for reliable interpretability tools has become pressing. Shapley values are a local explanations which gained popularity in recent years. Here, we reveal current misconceptions about the ``true to the data'' or ``true to the model'' trade-off and demonstrate its importance in a clinical context. We show that the interpretation of Shapley values, which strongly depends on the choice of a reference distribution for modelling feature removal, is often misunderstood. We further advocate that for applications in medicine, the reference distribution should be tailored to the underlying clinical question. Finally, we advise on the right reference distributions for specific medical use-cases. 
%In doing so, we expand on the established ``true to the data'' or ``true to the model'' trade-off for clinical settings. 
%However, feature removal modelling within SV forces the user to compromise between reliability and model faithfulness. Ultimately, SV are deemed unfit for explaining safety-critical models such as clinical ones.
\end{abstract}

\vspace*{-0.2cm}
\section{Introduction}
Since the introduction of KernelSHAP \cite{lundberg2017unified}, Shapley values have become standard in the model explainability literature. Originally proposed to attribute the effect that a player has on the outcome of a game, there have been some challenges in transferring this game theoretic notion to model interpretation. Given a model $\blackbox(\localobs_1,\ldots,\localobs_{\numfeats})$, the features from 1 to $\numfeats$ can be considered players in a game in which the payoff $\blackbox$ is a measure of the importance or influence of that subset. While Shapley values in game theory depend on the ordering of the players, Shapley values for model interpretation ``remove'' or drop features instead. When a feature is removed, its value is sampled from a reference distribution. When it isn't removed, its value is set to the feature value of our instance. The choice of a reference distribution remains challenging \cite{aas2019explaining, janzing2020feature, frye2020shapley}. %Please refer to \cite{covert2020explaining} for an overview of removal-based explanation methods. 
Recent papers on Shapley values have pointed out several limitations for each of the different distribution choices  \cite{kumar2020problems, fryer2021shapley}, causing the Shapley values to be seen as unstable and as such, unfit for safety-critical domains such as healthcare. By contrast, we argue that Shapley values are not inadequate but instead that the main challenges arise from the misconceptions around their approach to explanation. We advocate for a careful analysis of the possible reference distributions in light of the underlying research question before using Shapley values in a medical settings.
%or discussed the ``right'' distribution to simulate dropped features \cite{aas2019explaining, janzing2020feature, frye2020shapley}. 
\vspace*{-0.1cm}
\subsection{Contributions}
\begin{enumerate}[itemsep=0.5mm]
    \item We explicitly define the clinical questions Shapley values can respond to, for each reference distribution. Contrasting with \cite{chen2020true} who only differentiate between observational and interventional conditional reference distributions, we adapt our discussion to clinical contexts and take a broader viewpoint by integrating causality-based Shapley values.
    \item We use carefully chosen counter-examples to highlight the existing misinterpretations of Shapley values and illustrate how certain choices of reference distributions may lead to misleading results. By doing so, we go beyond the misconceptions around the ``true to the data'' or ``true to the model'' trade-off. We further show that Shapley values suffer from limitations that stem from model under-performance.
    \item We discuss which reference distribution is best suited for the following use-cases: counterfactual fairness, statistical fairness, explainability and feature selection; and illustrate our statements using clinical examples.
\end{enumerate}

%The paper is organized as follows. A short introduction to Shapley values is given in Section \ref{sec:shapley_values}. Choices of reference distributions are considered in Section \ref{sec:ref_distrib}. Use cases for each reference distribution are defined in Section \ref{sec:use_cases}, as well as counter-examples for misconceptions about these. Concluding remarks are provided in Section \ref{sec:discussion}.

\section{Background} 
\label{sec:shapley_values}

In model explainability, Shapley values quantify the contribution of various features $\{1,\ldots,\numfeats\}$ in the prediction of a complex model $\blackbox:\mathbb{R}^{\numfeats}\rightarrow\mathbb{R}^{\numout}$ at an instance $x$ as follows: $\blackbox(x)=\phi^{\blackbox}_{0}(x)+\sum_{i=1}^{M} \phi^{\blackbox}_{i}(x)$ where $\phi^{\blackbox}_{j}(x)$ is the Shapley value of feature $j$ to $\blackbox(x)$ and $\phi^{\blackbox}_{0}(x)=\mathbb{E} f(X))$ is the average prediction with the expectation over the observed data distribution. The contribution of a feature $j$ is computed based on the change in value function $\valuefct$ comparing when the feature $j$ is equal to the value $x_j$ with when it is removed from the input. To account for the dependence with other features, one takes the difference in value function $v$ averaged over all possible coalitions $\includedfeats$ of features excluding feature $j$. The value function is defined as the expectation of the black box algorithm at observation $\localobs$ over the not-included features $\droppedfeats$ using a reference probability distribution over feature values $\refdistvar{}{}$ such that $\valuefct(\feat{\includedfeats})=\expect_{\refdistvar{}{}}[\blackbox({\localobs_{\includedfeats}, \varimputed_{\droppedfeats}})]$ for $\droppedfeats:=\{1,\ldots,\numfeats\}/\includedfeats$ and the operation $({\localobs_{\includedfeats}, \localobs_{\droppedfeats}})$ denoting the concatenation of its two arguments. Ultimately, a binomial weight $\frac{{|\includedfeats|!(\numfeats-|\includedfeats|-1)!}}{m!}$ is used to recover the original Shapley values which account for all possible orderings. As a result, the Shapley values of feature $j$ is defined as follows:
\vspace*{-0.3cm}
\begin{align*}
    \phi_j^\blackbox(x) 
    &=   \frac{1}{\numfeats} \sum_{\substack{|\includedfeats|\in \\ \{0,\ldots,\numfeats-1\}}} \left[\frac{1}{\binom{\numfeats-1}{|\includedfeats|}} \sum_{\substack{\includedfeats\subseteq \{1,\ldots,\numfeats\}/j\; \\\text{with }|S|=j}} \Delta v_f(S, j, x)\right] \\
    &= \expect_{S}\left[\expect_{\refdistvar{}{\includedfeats}}[\blackbox({\localobs_{\includedfeats}, \varimputed_{\droppedfeats}})]\right]
\end{align*}
where $\Delta v_f(S, j, x) = v_f(S \cup \{j\}, x) - v_f(S, x)$

An equivalent approach to Shapley values is to consider that the attribution of feature $j$ equal to the expected change in the value function of including feature $j$ after a random number of features has already been included 
\vspace*{-0.3cm}
\begin{align*}
    \phi_{\valuefct}(j) 
    &=   \frac{1}{\numfeats} \sum_{\substack{|\includedfeats|\in \\ \{0,\ldots,\numfeats-1\}}} \left[\frac{1}{\binom{\numfeats-1}{|\includedfeats|}} \sum_{\substack{\includedfeats\subseteq \{1,\ldots,\numfeats\}/j\; \\\text{with }|S|=j}} \phi_{\valuefct}(j, \includedfeats)\right] \\
    &=  \expect_{|S|}\left[\expect_{S\given |S|}[\phi_{v}(j, S)]\right]. 
\end{align*}
This allows us to interpret the Shapley value of feature $j$ as the expectation of the following experiment: (1) sample the number of included features $|S|$ uniformly, (2) sample $|S|$ features from $\{1,...,m\}/j$ and (3) return the change in value function when including feature $j$ given $\includedfeats$ had been included.

Therefore, the Shapley value of feature $j$ answers the question: ``What is the {expected change} in the value function if we first evaluate it on a random subset of features not including $j$ before we evaluate it additionally on $j$?''. In practice, over the reference distirbution is estimated using a sample mean over a finite number of reference points. Further, the choice of a reference distribution to sample from has been subject to recent debates \cite{aas2019explaining, janzing2020feature, merrick2020explanation}. Marginal Shapley values \cite{lundberg2017unified, janzing2020feature} define $\refdistvar{}{\includedfeats}:=\dist({\varimputed})$ where $\dist$ denotes the marginal data distribution. Conditional Shapley values \cite{aas2019explaining} set the reference distribution equal to the conditional distribution given $\localobs_{\includedfeats}$ $\refdistvar{}{\includedfeats}:=\dist({\varimputed|\varimputed_{\includedfeats}=\localobs_{\includedfeats}})$. Since Shapley values are an expectation of $\valuefct(\includedfeats)$ over all coalitions. All in all, Shapley values for model interpretability are thus defined by 
\vspace*{-0.3cm}
\begin{align*}
    \phi(j) &= \expect_{S}\left[\expect_{\refdistvar{}{\includedfeats}}[\blackbox({\localobs_{\includedfeats}, \varimputed_{\droppedfeats}})]\right] \\
    & =\expect_{\refdistvar{}{}}\left[\phi_{\imputed}(j)\right]
\end{align*}

%where the second equality (denoted by $(*)$ above) only holds for marginal reference distributions.
%& \overset{(*)}{=} \expect_{\refdistvar{}{}}\left[\expect_{S}[\blackbox({\localobs_{\includedfeats}, \varimputed_{\droppedfeats}})]\right] \\

%When a feature is included in the coalition its value is set to the observed instance value $\localobs_{\includedfeats}$. When a feature is included in the coalition its value is sampled from a reference distribution $\refdistvar{}{\includedfeats}$. Ultimately, the value function is the following expectation $v_f(\feat{\includedfeats})=\expect_{\refdistvar{}{\includedfeats}}[\blackbox({\localobs_{\includedfeats}, \varimputed_{\droppedfeats}})] $ for $\droppedfeats:=\{1,\ldots,\numfeats\} \backslash \includedfeats$ with $({\localobs_{\includedfeats}, \localobs_{\droppedfeats}})$ denoting the concatenation of its two arguments. 

\section{Marginal, Conditional and Causality-based Shapley values}
\label{sec:ref_distrib}

\subsection{Marginal Shapley values}
Lundberg et. al \yrcite{lundberg2017unified} suggest using marginal Shapley values by sampling the dropped features $\droppedfeats$ from the data set for the computation of the value function $\valuefct({\includedfeats}) = \expect_{\refdistvar{\droppedfeats}{\includedfeats}}[\blackbox({\localobs_{\includedfeats}, \varimputed_{\droppedfeats}})]$. Janzing et. al \yrcite{janzing2020feature} show that the value function then becomes an expectation where the included features have been ``intervened on'', that is,  set to a value using the $do$ operation $\valuefct({\includedfeats}) = \expect[\blackbox(\varobs_{\includedfeats}, \varimputed_{\droppedfeats})\given \judeado{\left(\varobs_{\includedfeats}=\localobs_{\includedfeats}\right)]}.$ In this setting, the reference distribution is $\refdistvar{\droppedfeats}{\includedfeats}=\dist({\varimputed_{\droppedfeats}})$ where $\dist{}$ is the data distribution of $\varobs$. Consider a clinical model $\blackbox$ predicting the probability of a disease from patient features. The question the marginal Shapley value of feature $j$ for patient $\localobs$ answers is, ``What is the expected change in predicted disease probability if we were to set feature $j$ of a random patient $\imputed$ of the cohort equal to $\localobs_j$ after we already set a random subset of features (excluding $j$) equal to $\localobs$?''. 

Marginal Shapley values are often described as ``true to the model'' \cite{chen2020true}, which means that attributions depend on whether the features are explicitly used by the black-box. Although this is partly true, we contrast by showing that marginal Shapley values are still heavily influenced by the distribution of the data. 

Consider a linear model which predicts cardiovascular diseases $\blackbox(\localobs)=\localobs_1+\localobs_2$ where $\localobs_1$ is a measure of the patient's cholesterol relative to the recommended upper limit and $\localobs_2$ is a patient's fasting blood sugar (in g/L). We assume $\localobs_1\sim\Uniform(-1, 2)$ and $\localobs_2\sim\Uniform(0, 3)$ and a patient $\localobs=(0, 0)$. The Shapley value for $\localobs_1$ will be -0.5 and the Shapley value for $\localobs_2$ will be -1.5 (see Supplementary Material \ref{sec:suppl_marg_mean}). Here, the difference in magnitude between the two feature attributions results solely from the fact that they are differently far from their mean. In other words, it stems from the fact that the outcome of the model for our patient is differently far from the expected outcome when intervening on cholesterol compared to when intervening on fasting blood sugar value. Further, note that for the same black box model $\blackbox$ if $x_1$ and $x_2$ are centred on a common mean but have different spreads their two marginal Shapley values would not be equal, even though they play symmetric roles in the algebraic formulation of the model (see Supplementary Material \ref{sec:suppl_marg_spread}). Ultimately, for similar reasons if away from the center of mass, marginal Shapley values may lack locality \cite{ghalebikesabi2021locality}.\newline
Therefore, marginal Shapley values may still acknowledge the overall distribution of features and where our observation stands in this distribution, instead of depending solely on the model and the patient's features. In other words, the explanation given would be \emph{relative} to the rest of the cohort, and not faithful to the model only.  As model faithfulness is paramount in certain clinical contexts e.g. when evaluating the safety of a model, using marginal Shapley in these scenarios raises ethical concerns \cite{merrick2020explanation}. 

%To mitigate the dependence of the Shapley values on the data distribution, we could instead sample the different features from the same marginal distributions \cite{strumbelj2010efficient}. \cite{datta2016algorithmic} break the correlations between feature values in order to measure the influence of each feature value of interest on the outcome, independently of other correlated inputs. %Thus, while marginal Shapley values are in fact not solely true to the model, they behave exactly as expected.

\subsection{Conditional Shapley values}

Conditional Shapley values have value function $\valuefct({\includedfeats}) =\expect_{\dist({\varimputed_{\droppedfeats}}|{\localobs_{\includedfeats}})}[\blackbox(\varobs_{\includedfeats}, \varimputed_{\droppedfeats})\given \varobs_{\includedfeats}=\localobs_{\includedfeats}]$ where the reference function is the conditional distribution $\dist({\varimputed_{\droppedfeats}}|{\localobs_{\includedfeats}})$.
% $\refdistvar{\droppedfeats}{\includedfeats}=\dist({\varimputed_{\droppedfeats}}|{\localobs_{\includedfeats}}).$
For a clinical predictive model, these values answer the question, ``What is the expected difference in model outcome between a cohort with the same characteristics as our patient $\localobs$ for a random subset of features including $j$ and a cohort that has the same characteristics as our patient in the same random subset of features excluding $j$?''. 
Alternatively, \cite{covert2020explaining} show that using a conditional reference distribution can also be argued for from an information theoretic viewpoint: the model outcome of certain risk minimising black box models retrained on a restricted feature set can be approximated by the conditional expectation of the complete model outcome \cite{covert2020explaining}. From this view, conditional Shapley values answer the question, ``What is the expected change in a model's prediction assuming it was trained on a restricted feature set (excluding $j$) and we retrained it on the restricted feature set including $j$?''. 
Conditional Shapley values are often considered to be ``true to the data'', meaning that the model isn't evaluated on out-of-distribution instances i.e. implausible synthetic patients. Conditioning on the present features prevents from breaking the feature correlations observed in the reference population \cite{chen2020true, frye2020shapley, covert2020explaining}.
%Ultimately, we caution the reader against using a conditional reference distribution for cohorts with a high variety of covariate values across the cohort. A diversity in covariates induces that the model may generate misleading outcomes if evaluated off the data manifold.
%\cite{frye2020shapley} remark that conditionally imputed Shapley values converge to an explanation of how the information in the data associates with the labelled outcomes \cite{frye2020shapley}. Similarly \cite{covert2020explaining} say that conditional Shapley values provide 'well-grounded insight into intrinsic statistical relationships in the data'.

\subsection{Causality-based Shapley Values}
Interest for causal explanations has increased in the scientific community. Consequently, several approaches to causality based computations of Shapley values have been introduced in recent years. We refer to these as causality-based Shapley values. Such methods only consider causally-consistent orderings of single-reference games i.e. orderings such that known causal ancestors precede their descendants. End users are thus able to compute feature attributions even when they have partial knowledge of the causal graph underlying the data. These types of approaches favour explanations in terms of root causes, as they quantify the impact a given feature has on model prediction while its descendants remain unspecified. Asymmetric Shapley  \cite{frye2019asymmetric} uses conditioning by observation whereas Causal Shapley \cite{heskes2020causal} uses conditioning by intervention. Asymmetric Shapley values are said to be better for dataset with intrinsic ordering, such as temporal data, but Causal Shapley values are less sensitive to misspecifications of the causal graph structure \cite{heskes2020causal}. %This approach imputes the missing features with conditional distributions.
We notice that causality-based Shapley values strongly rely on the assumed graph. If the true causal structure deviates, the resulting explanation may be false. As such, causality-based Shapley values may ``force'' explanations to fit to the hypothesised causal flow. 
%Overall, causality-based Shapley Values may be more relevant in contexts where a specific treatment is of interest. If so, one needs to ensure this exposure isn't caused by the outcome itself e.g.~ a biomarker.

%statistical relationships between features can be highly misleading and uninformative about the true data distribution,

%\begin{comment}
%\begin{table}[H]
%\centering
%\resizebox{0.45\textwidth}{!}{%
%\begin{tabular}{@{}|l|c|@{}}
%\toprule
%Type of Shapley values & Value function $\valuefct({\includedfeats})$ \\ \midrule
%Marginal               &    $\expect_{\refdistvar{\droppedfeats}{\includedfeats}}[\blackbox({\localobs_{\includedfeats}, \varimputed_{\droppedfeats}})]$              \\
%Conditional            &      $\expect_{\dist({\varimputed_{\droppedfeats}}|{\localobs_{\includedfeats}})}[\blackbox(\varobs_{\includedfeats}, \varimputed_{\droppedfeats})\given \varobs_{\includedfeats}=\localobs_{\includedfeats}]$                  \\
%Causality-based        &                        \\
%Contrastive            &                        \\ \bottomrule
%\end{tabular}%
%}
%\caption{Summary of possible choices of reference distributions}
%\label{tab:ref_distrib}
%\end{table}

\section{Use cases for reference distributions}
\label{sec:use_cases}

\subsection{Counterfactual fairness}

One of the main benefits of marginal Shapley values is their dummy property, which says that only features that are in the model will get a non-zero attribution \cite{janzing2020feature, sundararajan2020many}. This property is of interest when examining the fairness of an algorithm. In particular, the notion of counterfactual fairness requires that for any individual, their model outcome remains unchanged if the \emph{sensitive} attribute (e.g. race, gender) had another value \cite{kusner2017counterfactual}.  
Counterfactual fairness is essential for any clinical application as using opaque decision making models may introduce or exacerbate disparities in care \cite{ahmad2020fairness}. Inspired by the example used from Merrick et. al \yrcite{merrick2020explanation}, we consider $\blackbox$ to be a black box algorithm for clinical management of a disease $\blackbox(\localobs)=\localobs_{male}$ with two binary features $\localobs_{male}$ and $\localobs_{eligible}$ where the latter indicates that a patient is eligible to surgery i.e.\ is likely to benefit from it. We further assume males are more likely to be eligible to surgery. Since intervening on the variable $\localobs_{eligible}$ and setting it to any arbitrary value will not have any direct effect on the model outcome, $\localobs_{eligible}$ receives a marginal Shapley value of 0. By contrast, $\localobs_{male}$ receives a marginal value of 1, as it is equal to the expected difference of the model outcome of the male population (here 1) and a population that has the same value as $\localobs{}$ either in $\localobs_{eligible}$ or not in any feature. \\
\\
\textbf{Proposition} \textit{Let $\blackbox$ be a black box model and feature $i$ be a sensitive attribute, having $\phi_i$=0, where $\phi_i$ denotes the marignal Shapley value of feature $i$, is a necessary, non sufficient condition to $\blackbox$ being a counterfactually fair w.r.t feature $i$.} 
\\
The proof of the above result is in the Supplementary Material \ref{sec:suppl_marg_fairness}, where we use an unfair function $\blackbox(\localobs)=\localobs_{male}\localobs_{eligible}$ in an all-male population and show the marginal attribution for gender is trivially 0. 
On the other hand, conditional Shapley values cannot be used to determine counterfactual fairness. In the previous examples, the conditional attribution of $\localobs_{eligible}$ is equal to a weighted sum of (a) the expected difference between the model outcome of the population which is eligible and the common population, and (b) the expected difference in model outcome between the population which is male and is eligible and the population which is male. Term (b) is null as both terms equal 1. Since the eligible sub-population will more likely be male than the common population, term (a) will be positive. Therefore, using conditional Shapley renders a positive attribution for $\localobs_{eligible}$ and the model will seem fairer than it is.

\subsection{Statistical fairness}

Statistical fairness covers criteria where we are interested in the expected difference in model outcome between the population with a specific sensitive attribute and the general population. These types of criteria -- such as having equal positive rates -- are popular in current medical AI practice as they tend to be easier to evaluate \cite{hedden2021statistical}. Frye et. al \yrcite{frye2020shapley} show that unfair models could be rephrased on unprotected features that are correlated with a sensitive attribute such that the outcome of the model remains the same for observations on the data manifold. Two models with different algebraic formulations but which evaluate the same on the data can thus render highly different marginal Shapley values whilst keeping close conditional Shapley values. This makes it challenging to judge the statistical fairness of an algorithm using a framework based on feature removal, especially in sensitive environments.

\subsection{Feature selection}
%Development of a severity of illness scoring system (inpatient triage, assessment and treatment) for resource-constrained hospitals in developing countries
When developing prediction models for medical decision making, feature selection is essential to build parsimonious models for clinicians to work with, particularly in resource-limited contexts. For instance, a risk score is a measure of clinical severity. It is often built directly from the feature attributions computed based on a prognosis model \cite{olson2013development, bernabe2016development}. Here, explanation models are used for selecting the most important variables in order to obtain a simple scoring system. 

The information theoretic equivalence regarding conditional Shapley values shown by Covert et. al \yrcite{covert2020explaining} entails implies these can be used for feature selection, as a single coalition Shapley value of 0 for feature $j$ reflects that there is no change in model outcome from including feature $j$ in the training process. We disagree with that interpretation, as having a Shapley value of 0 is still possible when including $j$ has a small effect in some coalitions, and a greater impact in others. Consider for instance the Shapley value of feature 2 in  $\blackbox(\localobs)=\indicate{\localobs_{1}>1}\cdot 3\localobs_{2}-\indicate{\localobs_{1}\leq 1}\cdot \localobs_{2}$ at $\localobs=(0.5, 0.5)$. Including feature 2 decreases the model outcome as we know that $\localobs_1=0.5$ but in expectation over feature 1, including feature 2 will increase the model outcome (see Supplementary Material \ref{sec:suppl_featureselect} for a proof).
Further, Covert et. al \yrcite{covert2020explaining} note that their interpretation only holds for a restricted set of black box functions. Assuming we only have black box access to the function, we would not know if the model under study exhibits the desired properties. Ultimately, the computational burden of Shapley values for feature selection is heavy, as it requires computing all instance-specific Shapley values in the global population.

\subsection{Justification} 

Explainability can refer to a variety of concepts \cite{adadi2018peeking}. In the following, we focus on situations where one (e.g.~a clinician) needs to \emph{justify} a decision. For such a purpose, using marginal Shapley values may be misleading. In the former clinical management example if the physician does not have access to the eligibility of a patient but to their weight, the latter might be used as a proxy measure of their eligibility to the surgery and have non-zero feature attribution. A common misconception about conditional Shapley values is that they would correctly assign a non-zero attribution to strength due to the fact that they are ``true to the data'' \cite{chen2020true}, contrasting with their marginal equivalent. We claim that such a description is reductive as it overlooks the fact that Shapley values explain the expected change in \textit{model} outcome. In other words, if the black box model does not capture the data dependence between the target variable and the covariates, for instance, because of mis-specification or low performance, even conditional Shapley values will not be true to the data. 

Ultimately, a clinical decision should be explained in a \emph{causal} manner \cite{tonekaboni2019clinicians}. Introducing causal knowledge into the explanation model should prevent spurious associations from being given importance. Therefore, we argue that for justifying a decision \cite{adadi2018peeking}, causality-based Shapley values \cite{frye2019asymmetric, heskes2020causal, wang2021shapley} should be used. However, existing methods strongly rely on causal assumptions. We thus caution against using them when domain knowledge is insufficient to build a causal graph, and further add that the black box to be explained should a high predictive performance. Ultimately, we claim that further work is needed in this space for providing robust causal interpretations.

\section{Concluding remarks}
\label{sec:discussion}
We give an overview of the existing misconceptions around Shapley values and explain how they can translate into potentially harmful explanations in a clinical context. We further define use cases for each reference distributions. To the best of our knowledge, this is the first attempt at discussing the right explanation technique for a targeted clinical question. Ultimately, we claim that further work should be done on explainable methods for models that are \emph{causal} themselves.

%Further, being true to the data does not necessarily mean that explanations give insight into the true feature interactions. An irrelevant variable could be introduced into our hiring algorithm, i.e. probability of getting prostate cancer. Assuming that this disease is positively correlated with strength, the expected model outcome should increase when restricting the common population to those with a probability of getting this cancer. The reason why such a variable should be assigned zero attribution is that it is not \emph{causal} for the model outcome or the predicted variable. Therefore, we argue that for explanation models to be able to justify a decision \cite{adadi2018peeking}, causal Shapley values \cite{frye2019asymmetric, heskes2020causal, wang2021shapley} should be used, and under the condition that the black box has a high predictive performance. %Building upon a specification of the causal structure, this kind of Shapley values attributes the change in model outcome not to statistical but causal relationships. %The expected difference in model outcome between a population with a random set of features (including $j$) that are equal to $\localobs$ and the population with a random set of features (excluding $j$) that are equal to $\localobs$ only to feature $j$ if the causal parent of feature $j$ has been intervened on. note that \cite{chen2020true} argue that we need conditional Shapley values when we want feature attributions that are 'true to the data' \cite{chen2020true}. 

\newpage
\bibliography{bib}
\bibliographystyle{icml2022}

%%%%%%%%%%%%%%%%%%%%%%%%%%%%%%%%%%%%%%%%%%%%%%%%%%%%%%%%%%%%%%%%%%%%%%%%%%%%%%%
%%%%%%%%%%%%%%%%%%%%%%%%%%%%%%%%%%%%%%%%%%%%%%%%%%%%%%%%%%%%%%%%%%%%%%%%%%%%%%%
% APPENDIX
%%%%%%%%%%%%%%%%%%%%%%%%%%%%%%%%%%%%%%%%%%%%%%%%%%%%%%%%%%%%%%%%%%%%%%%%%%%%%%%
%%%%%%%%%%%%%%%%%%%%%%%%%%%%%%%%%%%%%%%%%%%%%%%%%%%%%%%%%%%%%%%%%%%%%%%%%%%%%%%
\newpage
\appendix
\onecolumn

\section*{Supplementary Material}

\section{Marginal Shapley values} 

Marginal Shapley values are heavily influenced by the distribution of the data.

\subsection{Shifted distributions}
\label{sec:suppl_marg_mean}

First, let us consider the linear model $\blackbox(\localobs)=\localobs_1+\localobs_2$ with $\localobs_1\sim\Uniform(-1, 2)$ and $\localobs_2\sim\Uniform(0, 3)$ and local observation $\localobs=(0, 0)$. We show that although they play symmetric roles in the algebraic formulation of the black box model, their marginal values aren't equal: $\phi_{1}=-0.5$ and $\phi_{2}=-1.5$. 

\begin{align*}
    \phi_{1} &= \sum_{S\subseteq \{\emptyset, x_{2}\}} \frac{|S| !(|F|-|S|-1) !}{|F| !} \\
    &\left[f_{S \cup\{i\}}\left(x_{S \cup\{i\}}\right)-f_{S}\left(x_{S}\right)\right]\\
    & = \frac{1}{2} (f_{1}(x_{1})-f_{\emptyset}(\emptyset)) + \frac{1}{2} (f_{1, 2}(x_{1}, x_{2}) - f_{2}(x_{2}))\\
    & \overset{*}{=} \frac{1}{2}E[f(x_{1}, X_{2})] - \frac{1}{2}E[f(X_{1}, X_{2})] \\ &+ \frac{1}{2} E[f(x_{1}, x_{2})] - \frac{1}{2} E[f(X_{1}, x_{2})]\\
    & = \frac{1}{2}P(X_{2}=1) - \frac{1}{2}P(X_{2}=1|X_{1}=1)P(X_{1}=1) \\ &+ \frac{1}{2} - \frac{1}{2} P(X_{1}=1)\\
    & = \frac{1}{2}p - \frac{1}{2}p\cdot 1 + \frac{1}{2} - \frac{1}{2}\cdot 1 = 0
\end{align*}
where we used the definition $f(x_S)=E[f(x)|do\;x_S]$ from \cite{janzing2020feature} for marginal Shapley values in equation $*$.
\begin{align*}
    2\phi_1 & = \frac{1}{3}[\int_{0}^{3} x_2 d x_2 ] - \frac{1}{9} [\int_{0}^{3} \int_{-1}^{2} (x_1+x_2) d x_2 d x_1 ] \\& - \frac{1}{3}[\int_{-1}^{2} x_1 d x_1 ]\\
    & = \frac{1}{3}[\int_{0}^{3} x_2 d x_2 ] - \frac{1}{3} [\int_{-1}^{2} x_1 d d x_1 ] \\& -\frac{1}{3} [\int_{0}^{3} x_2 d x_2 ] - \frac{1}{3}[\int_{-1}^{2} x_1 d x_1 ]\\
    & = - \frac{2}{3} [\int_{-1}^{2} x_1 d x_1 ] \\
    & = - \frac{2}{3} [\int_{-1}^{2} x_1 d x_1 ] \\
    & = - \frac{2}{3} [\frac{4}{2} - \frac{(-1)^2}{2}] \\
\end{align*}
Thus
$$\phi_1 = \frac{-1}{2} $$

Symmetrically for $x_2$,
$$2\phi_2 = - \frac{2}{3} [\int_{0}^{3} x_2 d x_2 ]$$
and therefore
$$\phi_2 = - \frac{3}{2}$$

\subsection{Different spreads}
\label{sec:suppl_marg_spread}

Let us consider a black box $\blackbox(\localobs)=\localobs_1^2+\localobs_2^2$ with $\localobs_1\sim\Normal(0, 1)$ and $\localobs_2\sim\Normal(0, 10)$ and local observation $\localobs=(0, 0)$. While the first marginal Shapley value is -1, the second one is -100 as the expected change of model outcome is higher when intervening on the common population by setting $\localobs_2=0$ compared to setting $\localobs_1=0$. 

Similarly to the previous section:
\begin{align*}
    \phi_1 & = - \int_{-\infty}^{+\infty} \frac{x_{1}^2}{\sqrt{2\pi}} \exp{\frac{-x_{1}^2}2} d x_1 ] \\
    & =\frac{1}{\sqrt{2\pi}}[-x_{1} \mathrm{e}^{-\frac{x_{1}^{2}}{2}}-\int_{-\infty}^{+\infty}-\mathrm{e}^{-\frac{x_{1}^{2}}{2}} \mathrm{~d} x_1 \\
\end{align*}
Solving separately:
$$
\int-\mathrm{e}^{-\frac{x_1^{2}}{2}} \mathrm{~d} x
$$
We substitute $u=\frac{x_1}{\sqrt{2}} \longrightarrow \frac{\mathrm{d} u}{\mathrm{~d} x_1}=\frac{1}{\sqrt{2}}$ $\longrightarrow \mathrm{d} x_1=\sqrt{2} \mathrm{~d} u$ :
$$
=-\frac{\sqrt{\pi}}{\sqrt{2}} \int \frac{2 \mathrm{e}^{-u^{2}}}{\sqrt{\pi}} \mathrm{d} u
$$

We notice the Gaussian error function below: 
$$
\int \frac{2 \mathrm{e}^{-u^{2}}}{\sqrt{\pi}} \mathrm{d} u =\operatorname{erf}(u)
$$
We plug in solved integrals:
$$
\begin{aligned}
&-\frac{\sqrt{\pi}}{\sqrt{2}} \int \frac{2 \mathrm{e}^{-u^{2}}}{\sqrt{\pi}} \mathrm{d} u \\
&=-\frac{\sqrt{\pi} \operatorname{erf}(u)}{\sqrt{2}}
\end{aligned}
$$
We undo the substitution $u=\frac{x}{\sqrt{2}}$ :
$$
=-\frac{\sqrt{\pi} \operatorname{erf}\left(\frac{x}{\sqrt{2}}\right)}{\sqrt{2}}
$$
Ultimately: 
$$
\int_{-\infty}^{+\infty} \frac{x^{2} \mathrm{e}^{-\frac{x^{2}}{2}}}{\sqrt{2} \sqrt{\pi}} \mathrm{d} x =[\frac{\operatorname{erf}\left(\frac{x}{\sqrt{2}}\right)}{2}-\frac{x \mathrm{e}^{-\frac{x^{2}}{2}}}{\sqrt{2} \sqrt{\pi}}]_{-\infty}^{+\infty}=1 
$$\\
Symmetrically for $x_2$,
\begin{align*}
    \phi_2 & = - \int_{-\infty}^{+\infty} \frac{x_{1}^2}{10\sqrt{2\pi}} \exp{\frac{-x_{1}^2}{200}} d x_1 \\
    & = -100\\
\end{align*}

\section{Counterfactual Fairness} \label{sec:suppl_marg_fairness}
Because of the dummy property of marginal Shapley values we have that 
$$ \text{counterfactual fairness} \rightarrow \text{marginal Shapley value of 0}$$
for deterministic models. 
The back direction as we will see does not hold: 
Let our feature space comprise two binary variables $x_{canLift}$ and $x_{male}$ with 
\begin{align*}
    P(x_{male}=1) &= 1 \\
    P(x_{canLift}=1) &= p\\
    P(x_{canLift}=1|x_{male}=1) &= p
\end{align*}
where $p$ is an arbitrary probability.

Our black box algorithm is $f(x)=x_{male}\cdot x_{canLift}$ and the feature attribution of $x_{male}$ for $x_{male}=x_{canLift}=1$ can be computed as follows
\begin{align*}
    & \phi_{male} = \\&  \sum_{S\subseteq \{\emptyset, x_{canLift}\}} \frac{|S| !(|F|-|S|-1) !}{|F| !} [f_{S \cup\{i\}}\left(x_{S \cup\{i\}}\right) \\& -f_{S}\left(x_{S}\right) ]\\
    & = \frac{1}{2} (f_{male}(x_{male})-f_{\emptyset}(\emptyset)) \\& + \frac{1}{2} (f_{male, canLift}(x_{male}, x_{canLift}) - f_{canLift}(x_{canLift}))\\
    & \overset{*}{=} \frac{1}{2}E[f(x_{male}, X_{canLift})] - \frac{1}{2}E[f(X_{male}, X_{canLift})] \\& + \frac{1}{2} E[f(x_{male}, x_{canLift})] - \frac{1}{2} E[f(X_{male}, x_{canLift})]\\
    & = \frac{1}{2}P(X_{canLift}=1) + \frac{1}{2} - \frac{1}{2} P(X_{male}=1) \\& - \frac{1}{2}P(X_{canLift}=1|X_{male}=1)P(X_{male}=1) \\
    & = \frac{1}{2}p - \frac{1}{2}p\cdot 1 + \frac{1}{2} - \frac{1}{2}\cdot 1 = 0
\end{align*}
where we used the definition $f(x_S)=E[f(x)|do\;x_S]$ from \cite{janzing2020feature} for marginal Shapley values in equation $*$.

\section{Feature Selection} 
\label{sec:suppl_featureselect} 
Let our feature space comprise two independent binary variables $\varobs_{1}, \varobs_{2}\sim \Normal(1, 1)$.
Our black box algorithm is defined by $f(x)=\indicate{\localobs_{1}>1}\cdot 3\localobs_{2}-\indicate{\localobs_{1}\leq 1}\cdot \localobs_{2}$ and the conditional feature attribution of feature $2$ at $\localobs=(0.5, 0.5)$ can be computed as follows
\begin{align*}
    \phi_{2} &= \sum_{S\subseteq \{\emptyset, 1\}} \frac{|S| !(|F|-|S|-1) !}{|F| !}[f_{S \cup\{i\}}\left(x_{S \cup\{i\}}\right)
    \\& -f_{S}\left(x_{S}\right)]\\
    & = \frac{1}{2}(f_{2}(x_{2})-f_{\emptyset}(\emptyset)) + \frac{1}{2} (f_{1, 2}(x_{1}, x_{2}) - f_{1}(x_{1}))\\
    & \overset{*}{=} \frac{1}{2}E[f(X_{1}, x_{2}=0.5)] - \frac{1}{2}E[f(X_{1}, X_{2})] \\& + \frac{1}{2} E[f(x_{1}=0.5, x_{2}=0.5)] - \frac{1}{2} E[f(x_{1}=0.5, X_{2})]\\
    & = \frac{1}{2} (0.5\cdot 0.5\cdot 3 -0.5\cdot 0.5 ) - \frac{1}{2}E[0.5\cdot 3X_2-0.5X_2] \\& - \frac{1}{2} 0.5 - \frac{1}{2}E[-X_{2}] \\
    & = \frac{1}{4} - \frac{1}{2} - \frac{1}{4} + \frac{1}{2} = 0
    % & = \frac{1}{2} (0.5\cdot 0.5\cdot 3 -0.5\cdot 0.5 ) - \frac{1}{2}E[0.5\cdot 3X_2-0.5X_2] - \frac{1}{2} 0.5 - \frac{1}{2}E[-X_{2}] \\
    % & = \frac{2}{8} - \frac{1}{4}(a-1) - \frac{1}{4} + \frac{1}{2} \\
\end{align*}
where we used $f_S(x_S)=E[f(x_S,X_{\bar{S}})]$ in equation $*$.

\end{document}